\documentclass{icrc29}
\usepackage{unites2e}
\usepackage{graphicx,amssymb,amsmath,times}
\begin{document}
\title[OSETI]{Optical SETI with Imaging Cherenkov Telescopes}
\author[J. Holder et al.] { 
 J. Holder$^a$, P. Ashworth$^a$, S. LeBohec$^b$, H.J. Rose$^a$,
 T.C. Weekes$^c$\\
(a) School of Physics and Astronomy, University of Leeds, UK\\
(b) Dept. of Physics, University of Utah, USA\\
(c) Harvard-Smithsonian Center for Astrophysics, USA
}
\presenter{Presenter: J. Holder (jh@ast.leeds.ac.uk), \
uki-holder-J-abs3-og27-poster}

\maketitle

\begin{abstract}
The idea of searching for optical signals from extraterrestrial civilizations
has become increasingly popular over the last five years, with dedicated
projects at a number of observatories. The method relies on the detection of a
brief ($\sim$~few~ns), intense light pulse with fast photon
detectors. Ground-based gamma-ray telescopes such as the Whipple 10m,
providing a large mirror area and equipped with an array of photomultiplier
tubes (PMTs), are ideal instruments for this kind of observation if the
background of cosmic-ray events can be rejected. We report here on a method
for searching for optical SETI pulses, using background discrimination
techniques based on the image shape.
\end{abstract}
\vspace{-0.4cm}
\section{Introduction}
Schwartz and Townes \cite{Schwartz61} were the first to suggest the idea of
searching for optical wavelength signals from extraterrestrial
civilizations. The idea received little attention, with SETI projects choosing
instead to focus on radio searches. Advances in laser technology have
encouraged a re-evaluation of the potential of optical techniques for
SETI. The most promising method is to search for intense pulses of optical
photons from candidate star systems: Howard et al. \cite{Howard04} recently
noted that, using current technology (10m reflectors as the transmitting and
receiving apertures and a $3.7\U{MJ}$ pulsed laser source), a $3\U{ns}$
optical pulse could be produced which would be detectable at a distance of
$1000\U{ly}$, outshining starlight from the host system by a factor of $10^4$.

A number of dedicated projects attempting to detect such a signal are now
operating \cite{Howard04,Lampton00,Wright01}. Typically, these consist of a
$\sim1\U{m}$ diameter reflector which is used to observe candidate stars. The
collected photons are sent to a beam-splitter which redirects the signal to
$\sim2-3$ fast photo-detectors. The output pulse of the photo-detectors is
passed to discriminators, and the signal is recorded if all discriminators
trigger within a short time coincidence window.  Howard et al. \cite{Howard04}
recently published the results of a survey of 6176 stars made over five years
with the $1.6\U{m}$ Wyeth telescope at the Harvard/Smithsonian Oak Ridge
observatory. They were sensitive to pulses with intensities greater than
$\sim100\U{photons}\UU{m}{-2}$.

While diffraction limited resolution would be necessary for any potential
transmitter in order to keep the beam width small, the optical properties of
the receiver are less critical. Essentially all that is required is a large
'light bucket' with which to collect as many photons as possible, and fast
photo-detectors in order to discriminate the pulsed signal from the steady
background starlight and night-sky light. Similar requirements exist for
ground-based gamma-ray telescopes. Covault \cite{Covault01} has suggested
using the STACEE gamma-ray telescope for optical SETI studies. STACEE provides
a mirror area of $2370\UU{m}{-2}$, enabling the detection of pulses down to
$2\U{photons}\UU{m}{-2}$; however, optical SETI observations would require a
dedicated observing mode, and the discrimination of SETI signals from
cosmic-ray signals is not trivial for this type of experiment. Eichler and
Beskin \cite{Eichler01} have also considered the use of ground-based gamma-ray
telescopes for OSETI and note the importance of their relatively large
fields-of-view and the use of more than one telescope for coincidence-based
rejection techniques. Finally, the MAGIC collaboration have implemented a
scheme to monitor for OSETI pulses at the hardware trigger level \cite{Armada04}. In this
paper we present an analysis method for searching for optical pulses in
archival data taken with the Whipple 10m ground-based gamma-ray telescope.

\section{The Whipple 10m Telescope as an OSETI Detector}

Imaging atmospheric Cherenkov telescopes use arrays of hundreds of PMTs to
record the shape and orientation of the Cherenkov image in the focal plane due
to particle showers in the atmosphere. Since 2000, the Whipple telescope has
been equipped with a high resolution imaging camera of 379 PMTs with
$0.12^{\circ}$ spacing, giving a field of view of diameter $2.5^{\circ}$. The
PMT camera is covered by light collecting cones which remove the dead area
between PMTs. The telescope has a total mirror area of $90\UU{m}{2}$, mounted
on an altitude-azimuth positioner.

\begin{figure}[h]
  \begin{center}
    \begin{tabular}{cc}
      \includegraphics*[height=6cm]{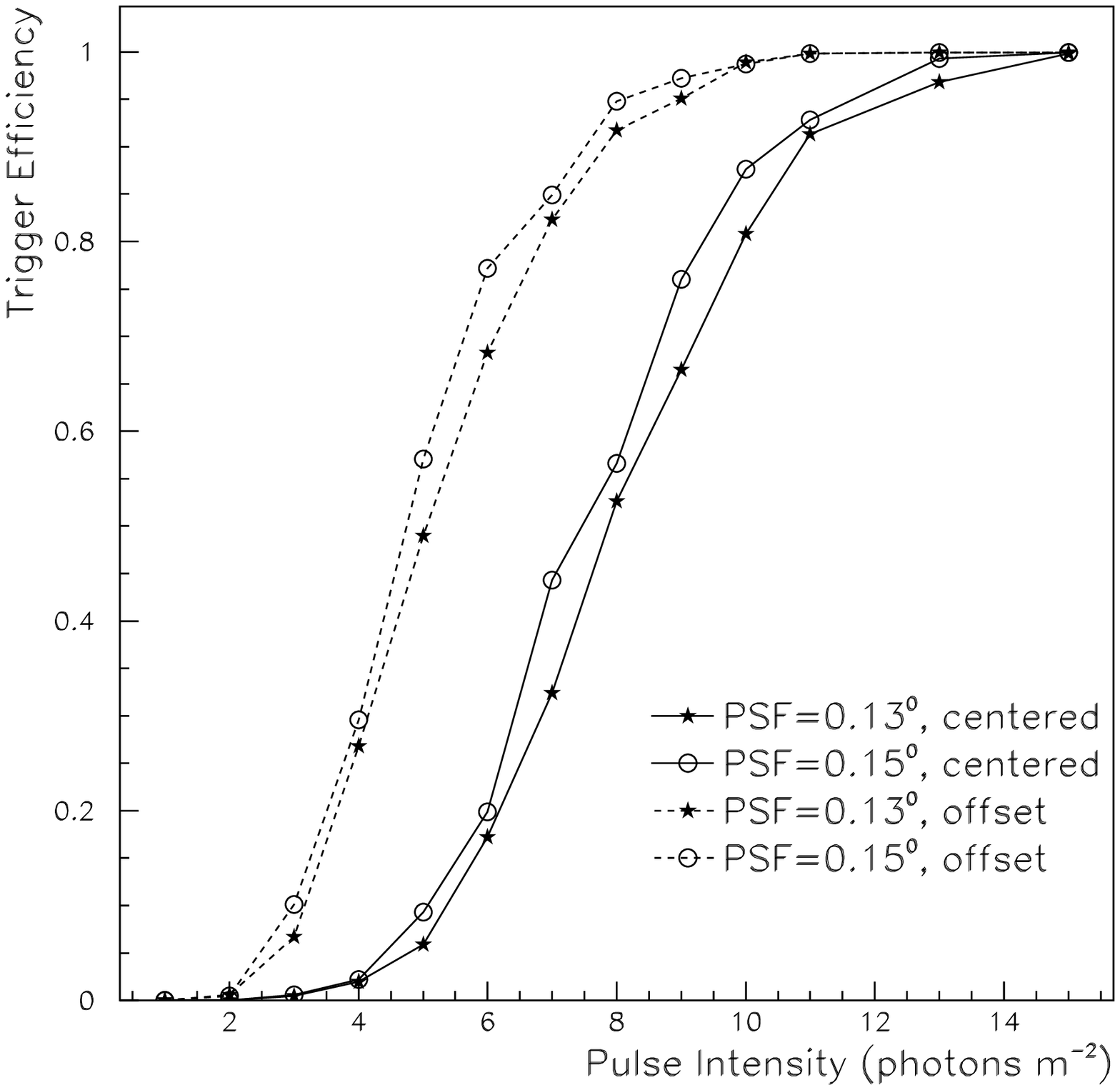}
      \hspace{1cm}
      \includegraphics*[height=6cm]{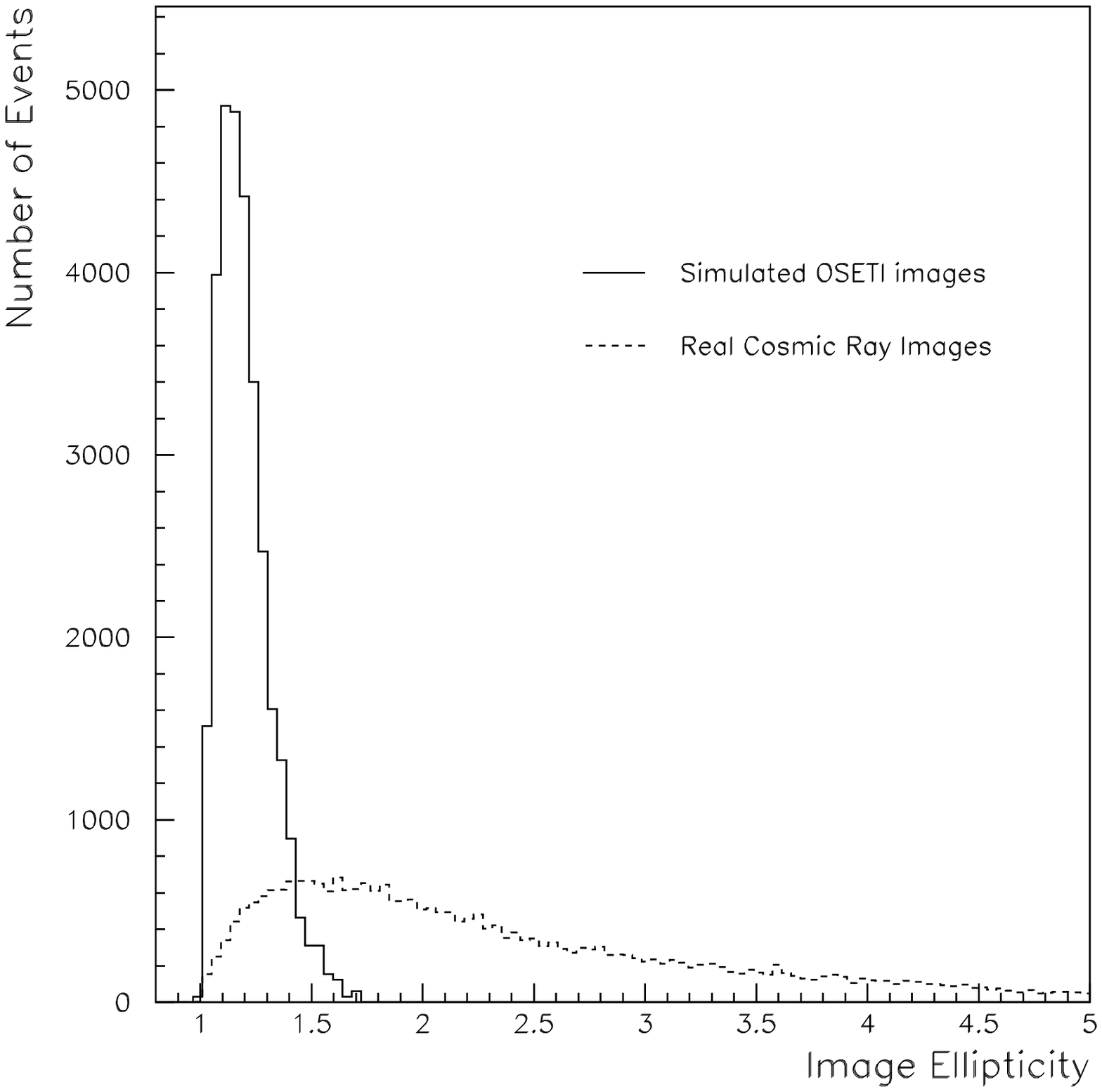}
    \end{tabular}
    \caption{\label {trig_eff} {\bf Left:} The simulated trigger efficiency as
  a function of pulse intensity. The simulations were run for two
  values of the telescope point spread function; $0.13^{\circ}$ and
  $0.15^{\circ}$. See text for details. {\bf
  Right:} The distributions of $ellipticity$ for simulated
  OSETI images and for real cosmic-ray  data. The OSETI images were
  generated with an intensity of $20\U{ph}\UU{m}{-2}$.  }
  \end{center}
\vspace{-0.3cm}
\end{figure}

An optical laser pulse coming from the direction of a candidate star will
appear as a point source in the camera of the Whipple telescope. If the optics
of the telescope were perfect, such a signal would never trigger the telescope
readout, since the photons would all fall within one PMT and the 3-PMT
multiplicity trigger condition would never be satisfied; however, the optical
requirements for Cherenkov telescopes are not strict, and the point spread
function (PSF) for the Whipple telecope is comparable to the PMT spacing. We
have performed a detailed simulation of the telescope response to point source
light pulses, including a complete simulation of the telescope optics and the
electronics chain \cite{grisu}. Figure~\ref{trig_eff} shows the trigger efficiency as a
function of pulse size for two values of the PSF and for two locations of the
pulse in the field-of-view; one at the centre of the central PMT
(corresponding to the worst case triggering scenario), the other offset to a
position equidistant from three PMTs close to the centre (best case). The
system has a trigger efficiency $>80\%$ for all cases for pulses of
$\sim10\U{ph}\UU{m}{-2}$, implying a sensitivity a factor of 10 greater than
that of the Harvard detector.

\vspace{-0.4cm}
\section{Analysis and Simulations}
The Whipple telescope is clearly able to trigger on low intensity,
point-source optical pulses; however, to provide a useful measurement we must
also be able to discriminate these pulses from the overwhelming background of
cosmic ray images. The image shape and orientation provide powerful
discrimination against cosmic rays for gamma-ray astronomy, and the same
techniques can be used to select point-source optical flashes. PMTs which
contain a signal above a threshold are selected, and the resulting image is
parameterized with an ellipse; the parameters of the ellipse
(\textit{length, width,} etc.) can then be used to identify candidate events.

\begin{figure}[h]
\vspace{-0.6cm}
  \begin{center}
    \begin{tabular}{cc}
      \includegraphics*[height=6cm]{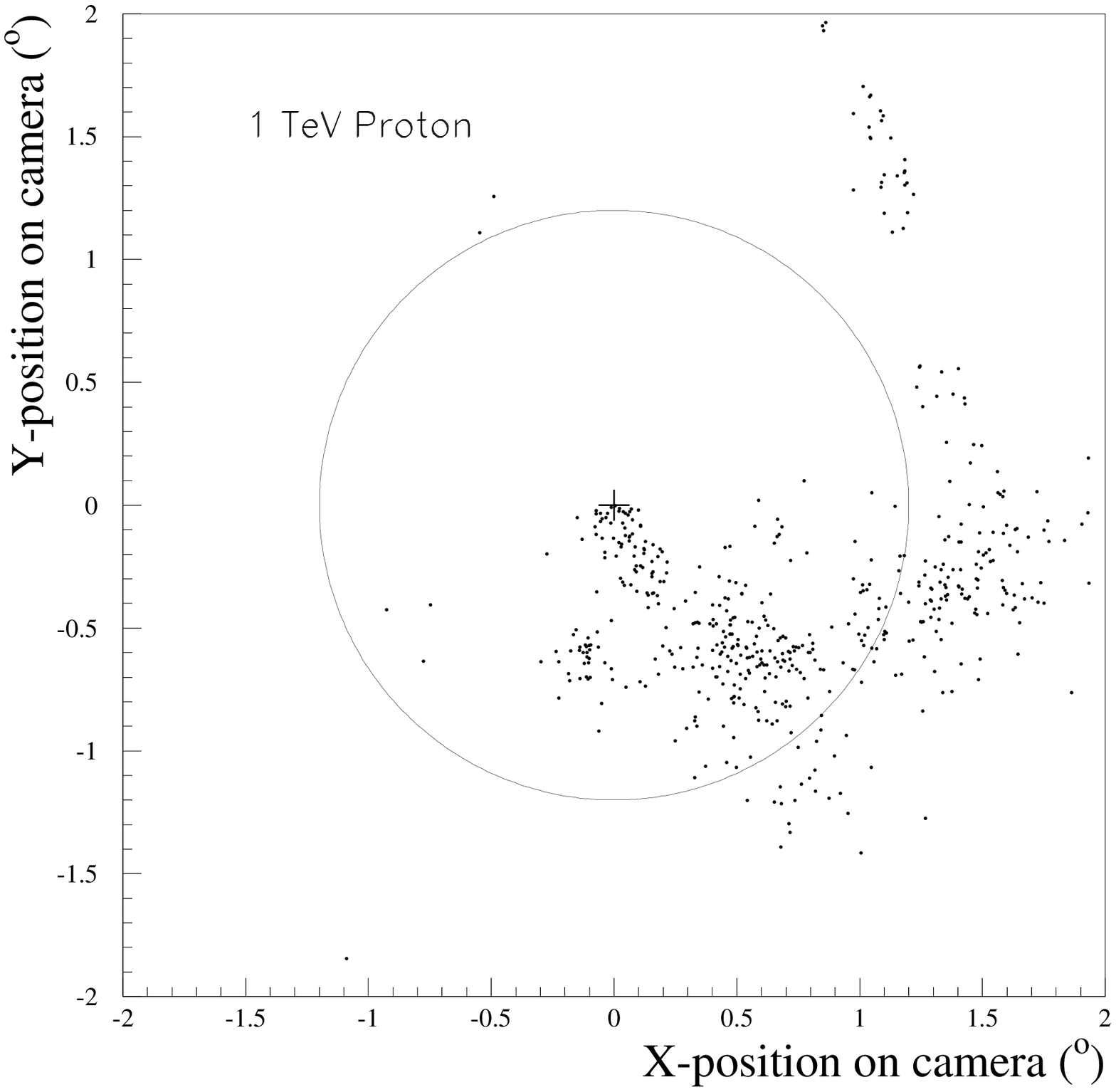}
      \hspace{1cm}
      \includegraphics*[height=6cm]{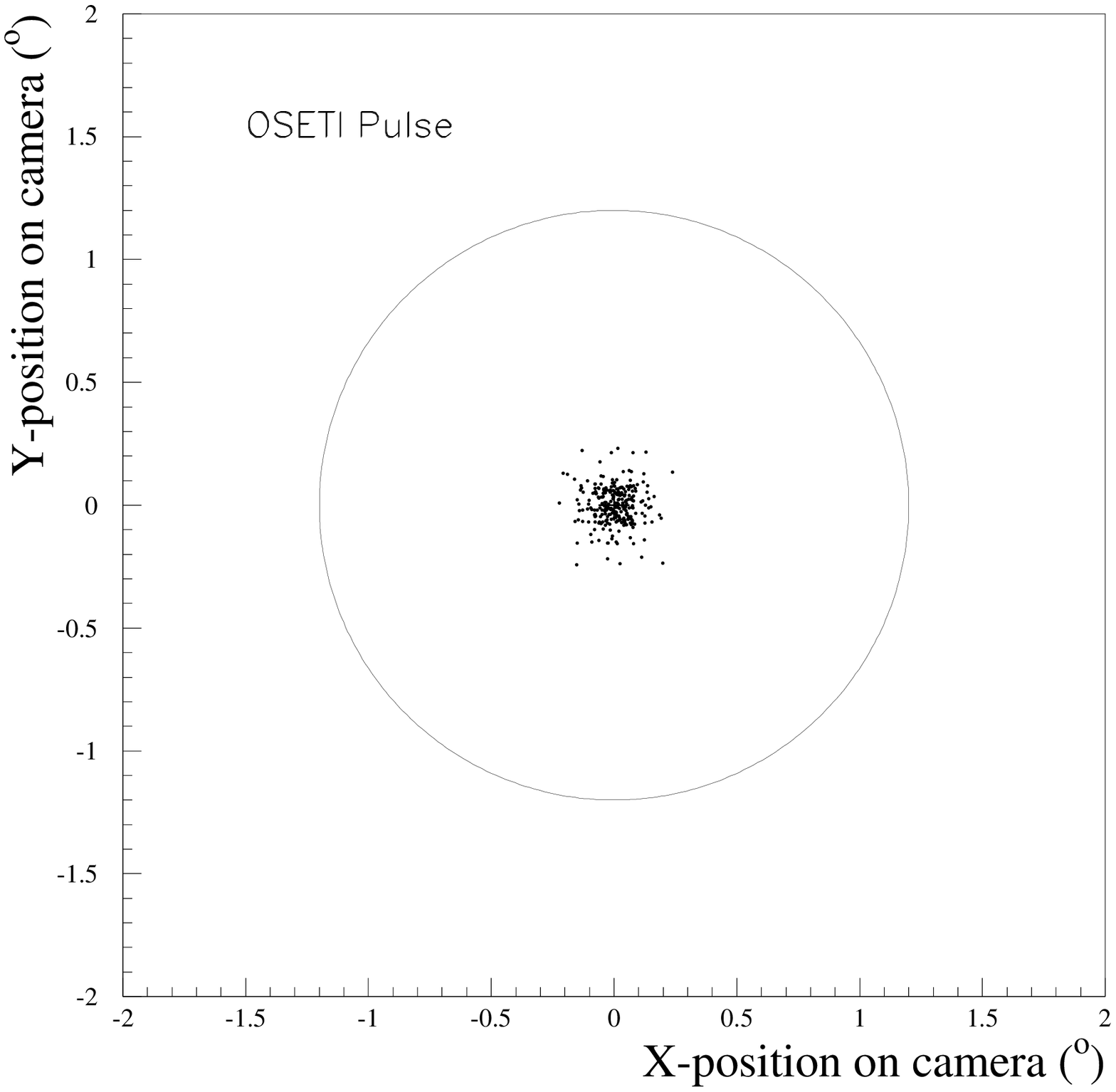}
    \end{tabular}
    \caption{\label {events}Photon positions in the focal plane for a
  simulated $1\U{TeV}$ proton cascade (left) and a point source laser flash
  with the source at the centre of the field-of-view (right). The large
  circle shows the extent of the Whipple 10m camera.  }
  \end{center}
\end{figure}

The image of a point source optical flash in the camera will be bright,
compact, symmetrical and co-located with the position of a candidate star, as
illustrated by the simulated OSETI flash shown in
Figure~\ref{events}. Figure~\ref{trig_eff} shows the $ellipticity$
(\textit{length/width}) of simulated OSETI flashes at the centre of the
camera, compared to the $ellipticity$ of cosmic rays for real data. Clearly,
the OSETI images are more symmetrical than the majority of the background
events; cutting events with ellipticity $>1.5$ reduces the background by 81\%,
while retaining 97.5\% of the OSETI images. Close to the edge of the camera
many cosmic-ray images become highly elliptical due to truncation. We
therefore restrict the field of view to images with an angular \textit
{distance} from the camera centre less than $1^{\circ}$.

\begin{figure}[h]
  \begin{center}
    \begin{tabular}{cc}
      \includegraphics*[height=6cm]{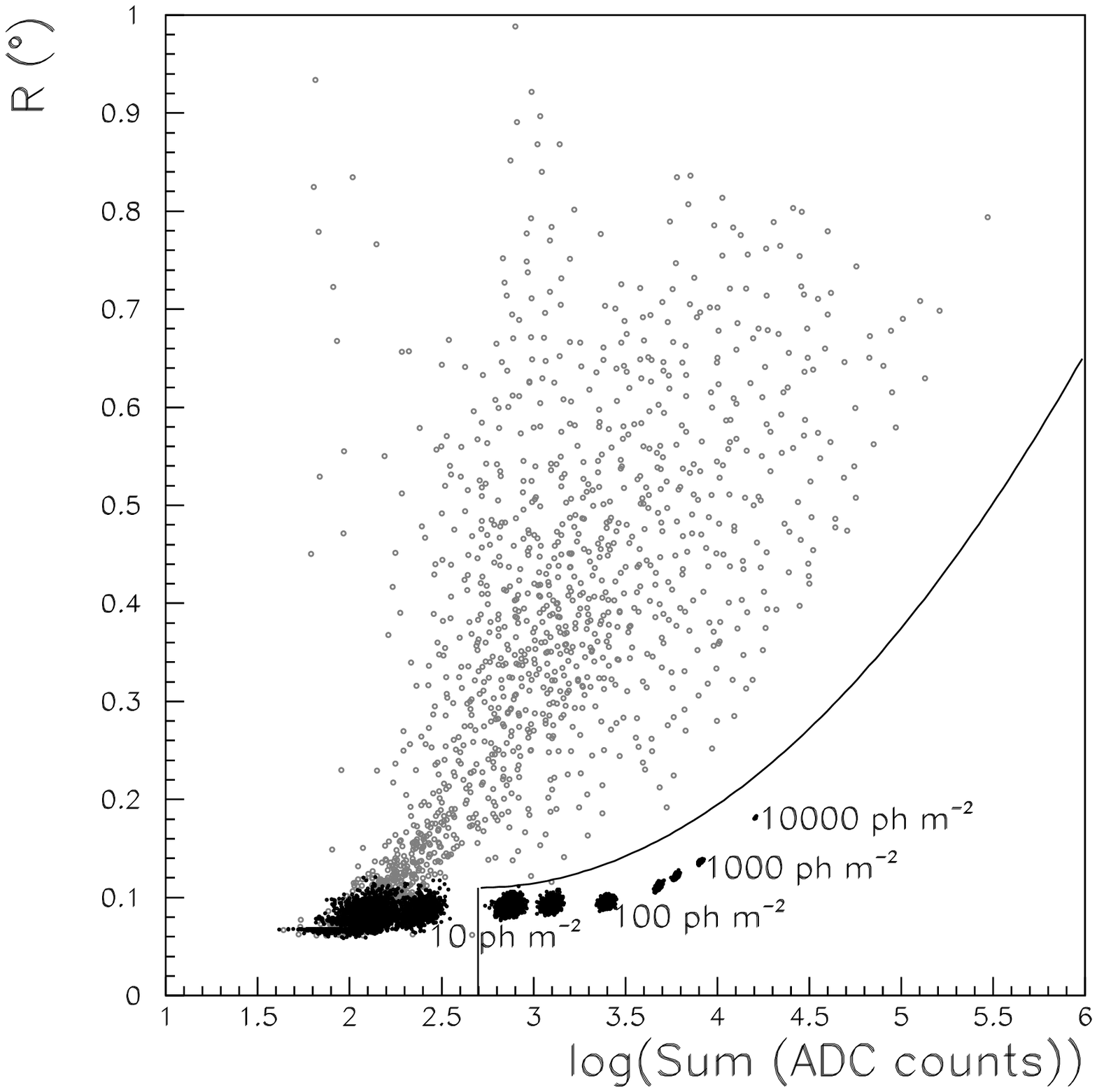}
      \hspace{1cm}
      \includegraphics*[height=6cm]{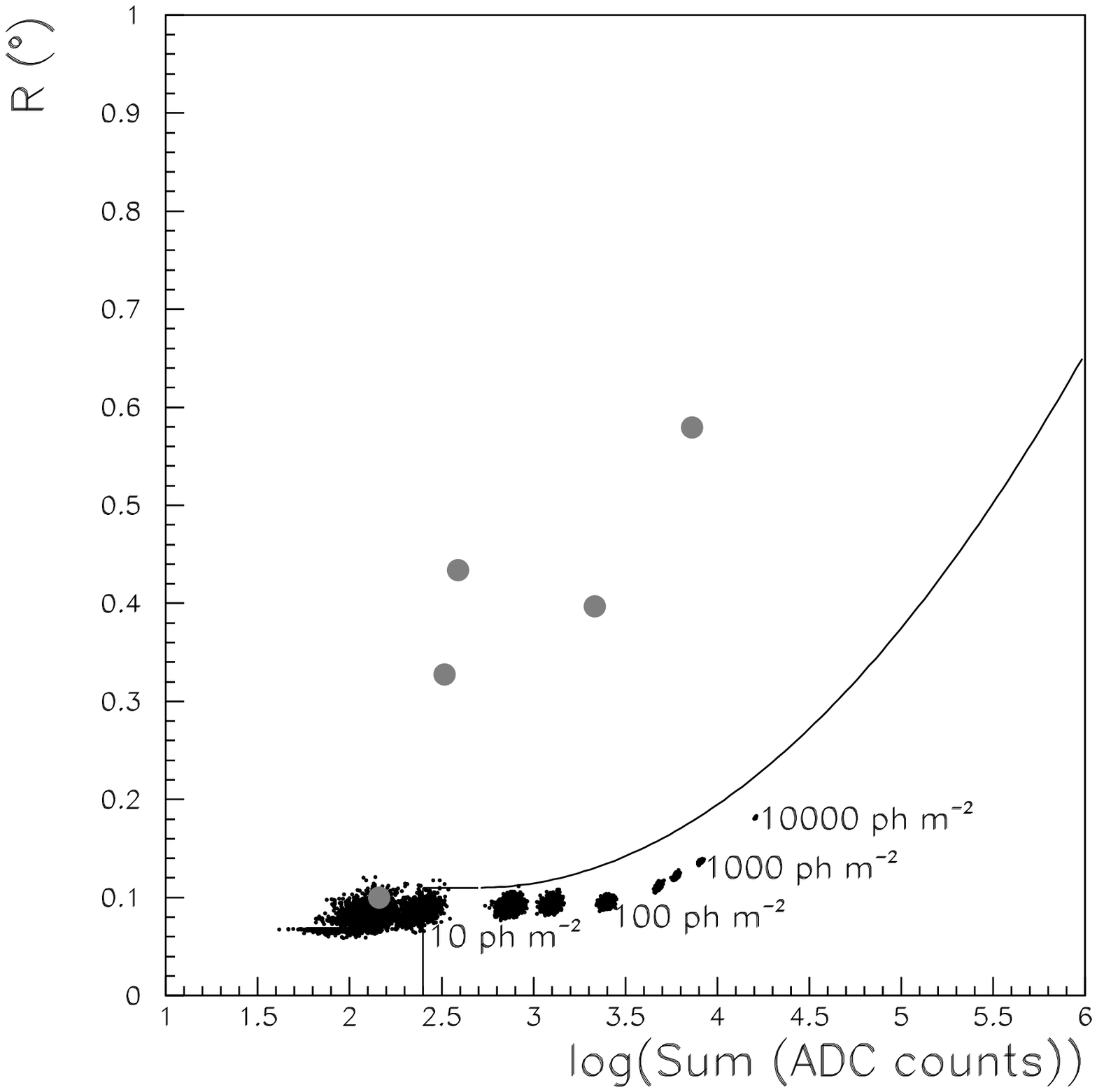}
    \end{tabular}
    \caption{\label {star}{\bf Left:} The image radius, R, as a function of
  the image charge /textit{sum} for simulated OSETI flashes (black points) and
  background cosmic rays (grey circles). The OSETI flash intensities at the
  telescope are indicated. The line describes the selection criteria for OSETI
  images (see text). {\bf Right:} The same for observations of HIP 107395,
  keeping only events from within $0.05^{\circ}$ of the source location at the
  centre of the camera.  }
  \end{center}
\vspace{-0.4cm}
\end{figure}

The most distinctive property of the OSETI flashes is their compactness. We
define the radius of the ellipse,
$R=\sqrt{width^2+length^2}$. Figure~\ref{star} shows $R$ as a function of the
$sum$ over all the charge in the image for cosmic rays and for OSETI
flashes simulated over a range of intensities, with the $ellipticity$ and
$distance$ cuts applied. Bright OSETI flashes can be clearly discriminated from the cosmic ray background; if
we select those events below and right of the line shown, only one of the
original 30900 cosmic-ray events remains.

The discrimination provided by the $ellipticity$, $distance$ and \textit{R vs
  sum} cuts is good; however, the signal strength is completely
  unknown and may consist of just one flash. Further discrimination is
  possible if we assume a candidate source position, such as the location of a
  star in the field of view. The statistical error on the centroid position of
  the image is small; $\sim0.01^{\circ}$ for flash intensities of
  $>10\U{ph}\UU{m}{-2}$; however the tracking errors of the telescope add a
  systematic error of $\sim0.05^{\circ}$. Selecting events which originate
  from within $0.05^{\circ}$ of a stellar candidate reduces the
  remaining background by a factor of $\sim400$.

As an example, we have made a targeted observation of one star, HIP 107395,
which was identified by Howard et al. \cite{Howard04} as their best candidate
for an OSETI pulse. The 28 minute observation was made on November
$\rm12^{th}$ 2004 beginning at 02:21 UTC. The right-hand side of figure
~\ref{star} shows the result. Only 5 events remain after the selection cuts on
$distance$, $ellipticity$ and image location. No events pass these cuts and
satisfy the \textit{R vs sum} selection, hence we have no evidence for optical
SETI signals from this candidate star during the observation, down to a
sensitivity limit of $10\U{ph}\UU{m}{-2}$.

\vspace{-0.4cm}
\section{Conclusions}
We have demonstrated that it is possible to search for optical SETI pulses
using atmospheric Cherenkov telescopes with a sensitivity an order of
magnitude better than dedicated experiments. While optical SETI observations
are unlikely to form a major part of the observing schedule for these
instruments, the large fields of view (from 2.5 to $5.0^{\circ}$) ensure that
candidate stars are visible during gamma-ray observations. The Whipple
telescope has been operating with an imaging camera for $>10$ years. A search
of this archive for OSETI pulses is currently underway. Other Cherenkov
experiments also have large data archives which could be analysed
similarly. The next generation of instruments (VERITAS, HESS, MAGIC, CANGAROO
III) provide a further increase in sensitivity, and background
rejection. Finally, we note that gamma-ray observations are not generally made
when the moon is visible to avoid damage to the PMTs. Targeted optical SETI
observations could be made during moontime with reduced sensitivity,
increasing the scientific output from these instruments.

\vspace{-0.5cm}
\section{Acknowledgements}
This research is supported by grants from the U.S. Department of Energy, the
National Science Foundation, the Smithsonian Institution, by NSERC in Canada,
Science Foundation Ireland and by PPARC in the UK. We thank the VERITAS
Collaboration for the use of their dataset.

\end{document}